\documentclass[twocolumn,showpacs,amsmath,amssymb,prb]{revtex4-1}

\usepackage{slashbox}
\usepackage{hhline}
\usepackage{bm}

\usepackage{graphicx}
\def\<{\langle}
\def\>{\rangle}
\def\({\left(}
\def\){\right)}
\def\[{\left[}
\def\]{\right]}
\def\up{\uparrow}
\def\dn{\downarrow}
\def\s{\sigma}
\def\e{\mathrm{e}}
\def\i{\mathrm{i}}

\def\Im#1{\mathrm{Im}\left\{#1\right\}}

\def\chib{\chi_0}
\def\orb#1#2#3#4{\mbox{{\scriptsize $\stackrel{#2}{#1} \,\stackrel{#3}{#4}$}}}
\def\d{\displaystyle}

\begin{document}
\title{Spin and Orbital Characters of Excitations in Iron Arsenide Superconductors Revealed by Simulated Fe $L$-Edge RIXS}

\author{E. Kaneshita$^{1}$}
\author{K. Tsutsui$^{2}$}
\author{T. Tohyama$^{3,4}$}
\affiliation{
$^1$Sendai National College of Technology, Sendai 989-8502, Japan\\
$^2$Synchrotron Radiation Research Center, Japan Atomic Energy Agency, Hyogo 679-5148, Japan\\
$^3$Yukawa Institute for Theoretical Physics, Kyoto University, Kyoto 606-8502, Japan\\
$^4$JST, Transformative Research-Project on Iron Pnictides (TRIP), Chiyoda, Tokyo 102-0075, Japan}

\date{\today}

\begin{abstract}
We theoretically examine the orbital excitations coupled to the spin degree of freedom in the parent state of the iron-arsenide superconductor, based on the calculation in a five-band itinerant model.
The calculated Fe $L_3$-edge resonant inelastic x-ray scattering (RIXS) spectra disclose the presence of spin-flip excitations involving several specific orbitals.
Magnon excitations predominantly composed of a single orbital component can be seen in experiments, although its spectral weight is smaller than spin-flipped interorbital high-energy excitations.
The detailed polarization and momentum dependence is also discussed with predictions for the experiments.
\end{abstract}

\maketitle

After intensive study of iron arsenides motivated by the discovery of high-temperature superconductivity, it has been recognized that both spin and orbital degrees of freedom are the key to understanding the physics of iron arsenides.

Spin excitations in the parent and superconducting compounds have been studied by inelastic neutron scattering experiments.~\cite{Lumsden}
In the parent compounds, anisotropic spin waves have been observed below the antiferromagnetic (AFM) transition temperature $T_\mathrm{N}$,~\cite{Diallo09,Zhao,Diallo10,Harriger} and the anisotropic spin excitations persist just above $T_\mathrm{N}$.~\cite{Diallo10,Ewings}
These behaviors are consistent with theoretical calculations within the random-phase approximation (RPA) for a five-orbital Hubbard model.~\cite{kaneshita}

In the AFM ordered phase with orthorhombic structure, an electronic anisotropy in the FeAs plane has been reported from scanning tunneling microscopy,~\cite{Chuang} optical conductivity,~\cite{Dusza,Nakajima} and angle-resolved photoemission spectroscopy (ARPES).~\cite{Wang,Kim}
Resistivity measurements have revealed an in-plane anisotropy not only in the orthorhombic phase but also in the tetragonal phase above the structural transition in BaFe$_2$As$_2$.~\cite{Tanatar,Chu}
This suggests a possible presence of a nematic order forerunning the structure transition and the magnetic ordering.
The ARPES measurements above $T_\mathrm{N}$ have shown the lift of degeneracy between $d_{zx}$ and $d_{yz}$ orbitals,~\cite{Yi} which indicates a strong influence of the orbital degree of freedom on the nematic order and the electronic anisotropy below $T_\mathrm{N}$.~\cite{CCChen}
In fact, the in-plane anisotropy of optical conductivity below $T_\mathrm{N}$ has been explained by taking into account orbital characters of interband excitations.~\cite{Sugimoto,ZPYin}

Since both spin and orbital characters contribute to the electronic states of iron arsenides, it is desired to detect both the excitations at the same time in the energy and momentum spaces.
For this purpose, we propose resonant inelastic x-ray scattering (RIXS) tuned for the Fe $L$ edge as a probe.
The $L$-edge RIXS consists of two processes:  the x-ray absorption process and the emission process by way of an intermediate state with core holes.
The x-ray absorption is accompanied by creation of an Fe $2p$ core hole and an electron in the $3d$ orbitals, where the $2p$ core hole state has a hybridized spin with the orbital angular momentum due to the spin-orbit coupling.
When the electron relaxed through the x-ray emission is not that excited from $2p$, the resulting orbital occupation of $3d$ electrons varies from the initial.

Note that the spin of the relaxed electron may be either up or down since the core hole spin is hybridized.
The final state may thus involves a spin flip.~\cite{deGroot,veenendaal,ament}
This means that we can investigate spin excitations in addition to orbital ones through the $L$-edge RIXS.
Recently, single magnon excitation has been observed in the Cu $L$-edge RIXS for cuprate superconductor La$_{2-x}$Sr$_x$CuO$_4$.~\cite{Braicovich}

In this study, we theoretically investigate magnon excitations and orbital excitations coupled to the spin degrees of freedom together with spectral features of Fe $L_3$-edge RIXS in iron arsenides.
Our calculations are performed for a five-band Hubbard model by using the RPA~\cite{kaneshita} and a fast-collision approximation.~\cite{veenendaal,Ament11}
In the AFM phase, we find that the magnon excitations predominantly composed of single orbital component appear in $L$-edge RIXS spectra with a weak intensity as compared with orbital excitations ---the orbital excitations lie just above those in contrast with the case of cuprates, where $d$-$d$ excitations are well separated from the magnon excitations.~\cite{Braicovich}
The dominant orbital excitations above the magnon mode are found to be accompanied by the spin-flip process, producing composite excitations of the coupled orbital-spin degrees of freedom.
The origin of the excitations is attributed to the spin-flip particle-hole excitation from occupied to unoccupied states.
We also predict the polarization and momentum dependence of the Fe $L_3$-edge RIXS prior to forthcoming experiments.

We calculate the ground state of a mean-field five-band Hubbard model written with the ordering vector $\mathbf{Q}=(\frac{2\pi}{N_Q},0)$ [$N_Q=2$ for AFM, and 1 for paramagnetic (PM) state]:
\begin{multline}
H_{MF}=\frac{1}{N_Q}\sum_{\mathbf{k},\s}\sum_{l,l'}\sum_{\mu,\nu}
\,d_{\mathbf{k}+l\mathbf{Q}\, \mu\, \s}^\dagger\, d_{\mathbf{k}+l'\mathbf{Q}\, \nu\, \s}
\\
\Bigg\{\bigg(
\sum_{\mathbf{\Delta}} t(\mathbf{\Delta};\mu,\nu)
\,\e^{\i (\mathbf{k}+l\mathbf{Q}) \cdot \mathbf{\Delta}}\
+\epsilon_{\mu}\,\delta_{\mu,\nu}\bigg)\,\delta_{l,l'}
\\
+\bigg[
-J\bigg(\sum_{\nu'}\<n_{(l-l')\mathbf{Q}\, \nu'\nu'\, \s}\>^*
-\<n_{(l-l')\mathbf{Q} \,\mu\mu\, \s}\>^*\bigg)
\delta_{\mu,\nu}
\\
+J
\(2\<n_{(l-l')\mathbf{Q} \,\mu\nu\, \s}\>^*
-\<n_{(l-l')\mathbf{Q} \,\nu\mu\, \s}\>^*\)
(1-\delta_{\mu,\nu})
\\
\hspace{1.2cm}
-U \<n_{(l-l')\mathbf{Q} \,\mu\nu\, \s}\>^*
\bigg]\,\(1-\delta_{l,l'}\)\Bigg\},
\label{eq:H}
\end{multline}
where $\mu$ and $\nu$ represent an electronic orbital, $\epsilon_\mu$ is the on-site energy, $t(\mathbf{\Delta};\mu,\nu)$ is the transition energy between Fe sites distanced by $\mathbf{\Delta}$, $d_{\mathbf{k}\,\mu\,\s}^\dagger$ is the creation operator of the $3d$ electron with wave vector $\mathbf{k}$ and spin $\s$ (quantized along the $x$ axis), $U$ and $J$ are the intraorbital Coulomb interaction and the Hund coupling, respectively, and the pair hopping is set equal to $J$.
We use the same parameter set as in Ref.~\onlinecite{kaneshita}: $U=1.2$ eV and $J=0.22$ eV, which yields the magnetic moment $0.8\mu_\mathrm{B}$.
The distribution of the local magnetization among different orbitals is as follows:
$3z^2-r^2$, $0.11\mu_\mathrm{B}$; $zx$, $0.13\mu_\mathrm{B}$; $yz$, $0.19\mu_\mathrm{B}$; $x^2-y^2$, $0.07\mu_\mathrm{B}$; $xy$, $0.29\mu_\mathrm{B}$.
The spin-density-wave order parameter is defined as
$\<n_{l\mathbf{Q}\,\mu\nu\,\s}\>=\frac{1}{N}\sum_{\mathbf{k}}
\<d_{\mathbf{k}+l\mathbf{Q}\,\mu\, \s}^\dagger\, d_{\mathbf{k}\,\nu\, \s}\>$
($N$ is the number of $\mathbf{k}$ points in the first Brillouin zone of the PM system, and $l\neq0$).

In terms of the quasiparticles diagonalizing $H_{MF}$, we consider particle-hole pair excitations on the mean-field ground state $|g\>$ with energy $E_g$ and write the Hamiltonian as $H_{\mathrm{ph}}$ for the particle-hole system.
We define the bare susceptibility as
${\chib^{s_1s_2}}_{\orb{\mu}{\nu}{\lambda}{\tau}}(\mathbf{q},\mathbf{q}',\omega)$
$=\frac{1}{N} \sum_{\mathbf{k}_0}\sum_{l,l'}\<g| d^\dagger_{\mathbf{k}_0+l'\mathbf{Q}\,\tau\,\sigma_2}
d_{\mathbf{k}_0+\mathbf{q}'+l'\mathbf{Q}\,\lambda\,\sigma'_2}$
$\frac{1}{\omega+i\eta-H_{\mathrm{ph}}+E_g} $
$d^\dagger_{\mathbf{k}_0+\mathbf{q}+l\mathbf{Q}\,\nu\,\sigma'_1}
d_{\mathbf{k}_0+l\mathbf{Q}\,\mu\,\sigma_1}|g\>$,
where $s_1=\up$, $\dn$, $+$, $-$ for the spin pair $(\s_1,\s'_1)=(\up,\up)$, $(\dn,\dn)$, $(\dn,\up)$, $(\up,\dn)$, respectively, and $s_2$ is associated with $(\s_2,\s'_2)$ as well.
We here set $\eta=0.01$ eV and $\mathbf{q}=\mathbf{q}'$.

We calculate the dynamical susceptibilities within the RPA
\begin{equation}
\hspace{-1cm}
\begin{pmatrix}
\chi^{+-}\\
\chi^{\up\up}\\
\chi^{\dn\up}\\
\end{pmatrix}
=
\begin{pmatrix}
\chib^{+-}\\
\chib^{\up\up}\\
0
\end{pmatrix}
-
\begin{pmatrix}
\chib^{+-} V^{-+} & 0 & 0\\
0&\chib^{\up\up}\,V^{\up\up}&\chib^{\up\up}\,V^{\up\dn}\\
0&\chib^{\dn\dn}\,V^{\dn\up}&\chib^{\dn\dn}\,V^{\dn\dn}
\end{pmatrix}
\begin{pmatrix}
\chi^{+-}\\
\chi^{\up\up}\\
\chi^{\dn\up}
\end{pmatrix},
\label{eq:chi}
\end{equation}
where the product of susceptibility and interaction between quasiparticles is taken as a matrix product represented in the orbital basis such as ${\d [\chib V\chi]_{\orb{\mu}{\nu}{\lambda}{\tau}} =\sum_{\mu'\nu'\lambda'\tau'}{\chib}_{\orb{\mu}{\nu}{\lambda'}{\tau'}} V_{\orb{\lambda'}{\tau'}{\nu'}{\mu'}}\chi_{\orb{\mu'}{\nu'}{\lambda}{\tau}}}$, and the nonzero elements of the interaction matrix are $V^{-+}_{\orb{\mu}{\mu}{\mu}{\mu}}=V^{\s\s'}_{\orb{\mu}{\mu}{\mu}{\mu}}=U$, $V^{-+}_{\orb{\nu}{\mu}{\mu}{\nu}}=U-2J$, $V^{-+}_{\orb{\nu}{\mu}{\nu}{\nu}} =V^{-+}_{\orb{\nu}{\mu}{\nu}{\mu}} =V^{\s\s'}_{\orb{\mu}{\mu}{\nu}{\nu}} =J$, $V^{\s\s'}_{\orb{\nu}{\mu}{\mu}{\nu}} =J-(U-2J)\delta_{\sigma,\sigma'}$, and $V^{\s\s'}_{\orb{\nu}{\mu}{\nu}{\mu}}=U-3J-J\delta_{\sigma,\sigma'}$ for $\mu\neq\nu$ ($\sigma$ and $\sigma'$ take $\up$ or $\dn$).
Below, we discuss the spin-transverse mode $\Im{-\sum \chi^{+-}}$, the spin-longitudinal mode $\Im{-\sum \(\chi^{\up\up}-\chi^{\dn\dn}\)}$, and the charge mode $\Im{-\sum \(\chi^{\up\up}+\chi^{\dn\dn}\)}$, where the summations are taken with respect to the four orbital indices.

In this study, we use $E_r$ as the unit of energy to evaluate the excitation energy under the presence of renormalization effects:
The larger the effects, the smaller the value of $E_r$, setting $E_r=1.0$ eV represents the case of no renormalization effects included.
Comparing our data of spin excitation with inelastic neutron scattering spectra,~\cite{Harriger,Ewings} we guess that the value of $E_r$ is around 0.5 eV.

\begin{figure}[t]
\begin{center}
\includegraphics[width = 0.7\linewidth]{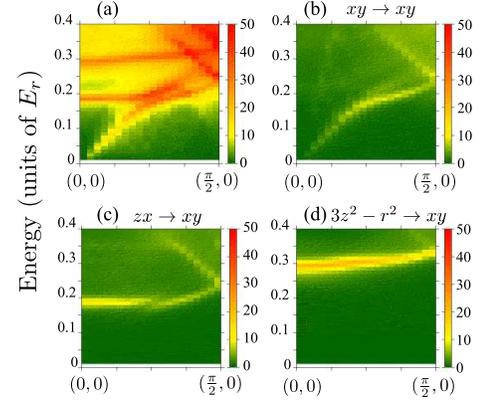}
\caption{(Color online) Spin excitation spectra of transverse mode for AFM case.
(a) the total spectrum is plotted.
Specific components corresponding to $\chi^{+-}_{\orb{\mu}{\nu}{\nu}{\mu}}$ is abstracted:
(b) $\mu=xy$ and $\nu=xy$,
(c) $\mu=zx$ and $\nu=xy$, and
(d) $\mu=3z^2-r^2$  and $\nu=xy$.
}
\label{fig:chi-trans}
\end{center}
\end{figure}
The calculated spectrum of the spin-transverse mode is plotted for the AFM case in Fig.~\ref{fig:chi-trans}(a):
The spectrum around (0,0) is shown here for RIXS study, although that around $(\pi,0)$ shows a strong intensity.
In the low-energy region ($<0.15E_r$), there is a sharp magnon branch stemming from (0,0).
Orbital analysis reveals that the magnon excitation is primarily associated with the $xy$-orbital component $\chi^{+-}_{\orb{xy}{xy}{xy}{xy}}$ as shown in Fig.~\ref{fig:chi-trans}(b).
This implies that the $xy$-orbital plays a significant role in the local magnetic moment, consistent with the calculated local magnetic moment, to which the $xy$ orbital component shows the largest (about 40\%) contribution while the others contribute less than 25\% each.

Above $0.15E_r$, there are flat excitation branches around $(0,0)$:
the $0.2E_r$ mode and $0.3E_r$ mode.
The orbital character of the $0.2E_r$ mode is a type of $zx \rightarrow xy$, which means that $\chi^{+-}_{\orb{zx}{xy}{xy}{zx}}$ is dominant [Fig.~\ref{fig:chi-trans}(c)].
On the other hand, that of the $0.3E_r$ mode is $3z^2-r^2 \rightarrow xy$ [Fig.~\ref{fig:chi-trans}(d)].

\begin{figure}[t]
\begin{center}
\includegraphics[width = 0.7\linewidth]{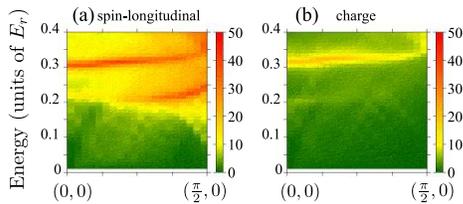}
\caption{(Color online) Excitation spectra of spin-longitudinal (a) and charge modes (b).
}
\label{fig:chi-longi}
\end{center}
\end{figure}
The spin-longitudinal and charge modes are plotted in Fig.~\ref{fig:chi-longi}.
Both show excitations around $0.3E_r$.
These $0.3E_r$ modes possess $3z^2-r^2 \rightarrow xy$ character as well as the spin-transverse mode.

The calculated density-of-states shows a peak in the $3z^2-r^2$ component of both the minority and majority spin states at the energy $\sim0.3E_r$ below the Fermi level and a rich $xy$ component of the majority spin just above the Fermi level illustrated in Ref.~\onlinecite{kaneshita}.
This coincides with the results here.
Thus, we conclude that the $0.3E_r$ mode in the AFM case simply arises from the interband transition.

In the spin-longitudinal mode, the flat excitation branch near $0.2E_r$ shows $zx \rightarrow xy$ character, and the dispersive part of this branch near $(\frac{\pi}{2},0)$ possesses $xy \rightarrow xy$ character as well.

\begin{figure}[t]
\begin{center}
\includegraphics[width = 0.4\linewidth]{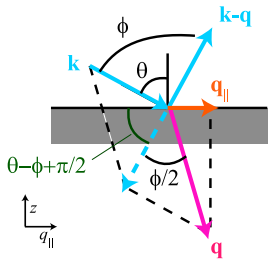}
\caption{(Color online) Situation setting.  Here, x-ray incoming with momentum $\mathbf{k}$ by angle $\theta$ is scattered to the angle $\phi$ ($=\frac{\pi}{2}$) with $\mathbf{k}-\mathbf{q}$ ($|\mathbf{k}-\mathbf{\mathbf{q}}|\approx|\mathbf{k}|$).
The incoming beam is polarized either perpendicular or parallel to scattering plane, and the scattered is a summation of both.
The $\mathbf{q}_{\parallel}$ axis is taken to the same direction as the in-pane component of $\mathbf{k}$; i.e., the illustrated $\mathbf{q}_{\parallel}$ is pointing the positive direction.
} \label{fig:setting}
\end{center}
\end{figure}
The excitation properties discussed above are expected to be observed by Fe $L$-edge RIXS experiments.
In Fig.~\ref{fig:setting}, we illustrate the geometry of our RIXS calculation.
The momenta of incoming ($\mathbf{k}$) and scattered ($\mathbf{k}'$) x-ray are perpendicular to each other $(\mathbf{k}\perp\mathbf{k}')$, and these norms are almost the same $k\approx k'\approx0.4526\pi$ in units of inverse lattice constant, where the lattice constant is assumed as 3.996~\AA.~\cite{chen}
The relation between the incident angle $\theta$ and the momentum transfer $\mathbf{q}=\mathbf{k}-\mathbf{k}'$ is determined as
$\theta = \frac{\phi}{2} + \arcsin{ \frac{q_{\parallel}}{q} }$
with $\phi$ ($=\frac{\pi}{2}$), the angle between $-\mathbf{k}$ and $\mathbf{k}'$.
The incoming beam is either $\sigma$ polarized (perpendicular to the scattering plane) or $\pi$ polarized (perpendicular to the $\sigma$ polarization), and the scattered is a summation of both.

The $L$-edge RIXS spectrum is calculated as a second-order perturbation involving the x-ray absorption and emission processes.
The transition operator $D_\mathbf{k}$ for the electron system in the absorption process is given, within the dipole approximation, as
$D_{\mathbf{k}} \approx \sum_{j, j_z,\mu, \sigma,\mathbf{k}'}
c^{j\,j_z}_{\mu,\sigma}(\bm{\varepsilon})
\, d^{\dagger}_{\mathbf{k}',\mu,\sigma}p_{\mathbf{k}+\mathbf{k}',j,j_z}
+\mathrm{H.c.},$
where $p$ is the annihilation operator of the Fe $2p$ electron, and the coefficient is the dipole transition matrix
$c^{j\,j_z}_{\mu,\sigma}(\bm{\varepsilon})
=\langle 3d;\mu,\sigma |\mbox{$\bm{\varepsilon}\cdot\mathbf{r}$}
| 2p;j,j_z \rangle,$
where $j$ and $j_z$ represent the total angular momentum and its $z$ component of the $2p$ electrons, respectively; $\bm{\varepsilon}$ is the unit vector of the polarization of the incoming or outgoing x-ray.

Assuming that the scattering occurs very fast, we employ the fast-collision approximation,~\cite{veenendaal,Ament11} which neglects the energy dispersion of the intermediate state.
The spectrum can be yielded from the calculated susceptibility and dipole transition matrix as
\begin{multline}
I_{RIXS}(\mathbf{q}_\parallel,\omega)\propto -\mathrm{Im}\Big\{
\sum_{\{\s_i\}} \sum_{\orb{\mu}{\nu}{\lambda}{\tau}}
\chi_{\orb{\mu}{\nu}{\lambda}{\tau}}^{s_1s_2}
(\mathbf{q}_\parallel,\omega)
\\
\times\Big[\sum_{{j_z,j'_z}}
c^{j\,j_z}_{\mu,\s_1}(\bm{\varepsilon}_\mathrm{o})^*
c^{j\,j_z}_{\nu,\s'_1}(\bm{\varepsilon}_\mathrm{i})
c^{j\,j'_z}_{\lambda,\s_2}(\bm{\varepsilon}_\mathrm{i})^* c^{j\,j'_z}_{\tau,\s'_2}(\bm{\varepsilon}_\mathrm{o})\Big]
\Big\},
\label{eq:I-RIXS}
\end{multline}
where the subscript i (o) of $\bm{\varepsilon}$ denotes incoming (outgoing) x ray; $s_1$ and $s_2$ are associated with $(\s_1,\s'_1)$ and $(\s_2,\s'_2)$, respectively; and we here set $j=3/2$, considering the Fe $L_3$-edge absorption.~\cite{j=1/2}
Due to the spin-orbit coupling of the Fe $2p$ electrons in the intermediate state, a spin flip can be achieved in the final state of the RIXS process ---the dipole transition matrix elements between the opposite spin states give finite contribution to the spectral intensity.
This allows us to analyze the spin-transverse mode apart from the charge excitation.

\begin{figure}[t]
\begin{center}
\includegraphics[width = 1.0\linewidth]{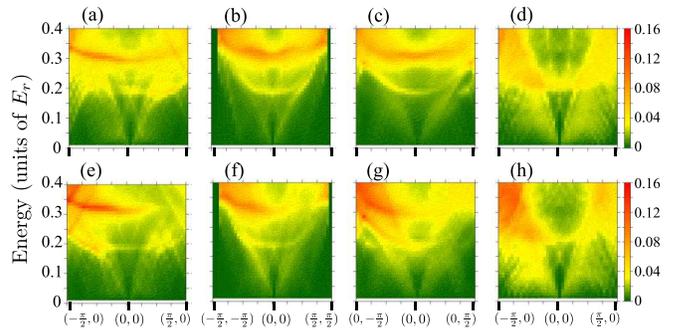}
\caption{(Color online) Fe $L_3$-edge RIXS spectra.
Spectra for AFM case are plotted along ($-\frac{\pi}{2}$,0)-($\frac{\pi}{2}$,0) in (a) and (e),  ($-\frac{\pi}{2}$,$-\frac{\pi}{2}$)-($\frac{\pi}{2}$,$\frac{\pi}{2}$) in (b) and (f), and  (0,$-\frac{\pi}{2}$)-(0,$\frac{\pi}{2}$) in (c) and (g); for the PM case, spectra are plotted along ($-\frac{\pi}{2}$,0)-($\frac{\pi}{2}$,0) in (d) and (h).
Polarization is set as $\sigma$ for the upper panels and $\pi$ for the lower.
} \label{fig:RIXS-AFM-PM}
\end{center}
\end{figure}
The calculated spectra for the AFM case are plotted in Figs.~\ref{fig:RIXS-AFM-PM}(a)--(c) and (e)--(g) along different directions of $\mathbf{q}_{\parallel}$ for different polarizations.
The fundamental structure of the spectra reflects the excitations discussed above, with the spectral intensity modified.
What kind of excitation is enhanced depends on the polarization and $\mathbf{q}_{\parallel}$ through the coupling of the spin and orbital degrees of freedom.
The magnon branch appears with a somewhat weak intensity, while the $0.3E_r$ mode appears still clear.

Note that the magnon observation in the present system is under the same condition as that in cuprates discussed in Ref.~\onlinecite{ament}:
In cuprates, the $x^2-y^2$ spin can flip in the $L$-edge RIXS process because the local spin moments lie in the CuO$_2$ plane.
In iron arsenides, the local magnetic moments also lie in the FeAs plane, where the dominant component, $xy$ orbital, forms another linear combination of states with the $z$ component of the orbital angular momentum $l_z = \pm2$.
Thus, the magnon in iron arsenides is also able to excite in the $L$-edge RIXS process.

We here describe the details of $\mathbf{q}_{\parallel}$ dependence and polarization dependence.
The magnon branch is rather visible along the diagonal path [Fig.~\ref{fig:RIXS-AFM-PM}(b) and (f)]:  better in the $+$/$-$ direction for the $\sigma$/$\pi$ polarization.
The $0.3E_r$ mode tends to be more distinct for the negative $\mathbf{q}_{\parallel}$.
This mode along ($-\frac{\pi}{2}$,0)-($\frac{\pi}{2}$,0) shows a spin-transverse feature (spin-flip) for the $\sigma$-polarization case, and shows spin-longitudinal and charge features (no spin-flip) for the $\pi$-polarization case.

In RIXS experiments using a twinned sample, the spectra will be obtained as a superposition of $x$ and $y$ directions.
In the spectra along (0,$-\frac{\pi}{2}$)-(0,$\frac{\pi}{2}$), the basic structure of the excitation spectra is similar to that along ($-\frac{\pi}{2}$,0)-($\frac{\pi}{2}$,0):  magnon, $0.3E_r$ mode, and $0.2E_r$ mode.
A difference appears near ($-\frac{\pi}{2}$,0)/(0,$-\frac{\pi}{2}$), e.g., around $0.2E_r$ [see Figs.~\ref{fig:RIXS-AFM-PM} (a) and (c)]:
this anisotropy arises from the $yz$/$zx$ orbital difference ---these excitations near ($-\frac{\pi}{2}$,0) show $yz \to yz$ character.
The spin character of $0.3E_r$ is also different.
For example, the spin-longitudinal (charge) component contributing to the $0.3E_r$ mode in the negative (positive) $\mathbf{q}_{\parallel}$ makes the spin-transverse component no longer dominant in the spectrum for $\sigma$ polarization.
For $\pi$ polarization, on the other hand, the spin-longitudinal component does not exist in the $0.3E_r$ mode.
To clarify the spin and orbital characteristics, RIXS experiments are required to be performed under the detwinned condition.

The spectra for the PM case are plotted in Fig.~\ref{fig:RIXS-AFM-PM}(d) and (h).
There are no distinct excitation structures near $0.3E_r$, in contrast to the AFM case.
$3z^2-r^2 \rightarrow xy$ excitations lie near $0.2E_r$ in the PM case; this orbital character is consistent with the calculated density of states (not shown).
The low-energy appearance is also different from the AFM case, spreading in the entire region.
The $\mathbf{q}_{\parallel}$ dependence of the spectral intensity appears in $\pi$ polarization.

In summary, we have investigated the spin and orbital characteristics of excitations in iron arsenides with a mean-field five-band Hamiltonian by RPA.
In addition, we have calculated the $L_3$-edge RIXS spectra with a fast-collision approximation.
The spin character to be observed becomes selective by making use of the momentum transfer dependence and the polarization dependence.
We have discovered the spin and orbital characteristics expected to be observed for certain conditions as summarized below.

We have reached a thorough understanding of the orbital characteristics of the magnon excitation in the AFM state, revealed that the magnon excitation mainly involves the $xy$ orbital, and concluded that the $xy$ orbital plays a key role in the local magnetic moment in both the static and the dynamical spin features.
We have confirmed that the magnon branch appears in $L$-edge RIXS spectra.
Since it is weaker than orbital excitations in the RIXS spectra, it is useful to observe along (0,0)-($\pm\frac{\pi}{2}$,$\pm\frac{\pi}{2}$), where $+$/$-$ is for the $\sigma$/$\pi$ polarization.

As for orbital excitations, a distinct branch has been found around $0.3E_r$ in the AFM case, determined as a particle-hole excitation from $3z^2-r^2$ to $xy$.
The difference between the AFM and PM states should be observed as the absence of this mode.
We propose an $L$-edge RIXS experiment under a detwinned condition to selectively observe the spin character of this mode.
It will show a spin-flip feature for $\sigma$ polarization but no spin-flip for $\pi$ polarization when observed along ($-\frac{\pi}{2}$,0)-(0,0).

We thank K. Zhou, R. A. Ewings, and K Sugimoto for fruitful discussions.
This work was supported by the Grant-in-Aid for Scientific Research from the MEXT of Japan; the Global COE Program ``The Next Generation of Physics, Spun from University and Emergence"; and Yukawa Institutional Program for Quark-Hadron Science at YITP.
A part of the numerical computation in this work was carried out at the Yukawa Institute Computer Facility.



\begin{thebibliography}{99}


\bibitem{Lumsden} For a review, see M. D. Lumsden and A. D. Christianson,
J. Phys.: Condens. Matter \textbf{22}, 203203 (2010).

\bibitem{Diallo09} S. O. Diallo \textit{et al.},
Phys. Rev. Lett. \textbf{102}, 187206 (2009).

\bibitem{Zhao} J. Zhao \textit{et al.},
Nat. Phys. \textbf{5}, 555 (2009).

\bibitem{Diallo10} S.  O. Diallo \textit{et al.},
Phys. Rev. B \textbf{81}, 214407 (2010)

\bibitem{Harriger} L. W. Harriger \textit{et al.}
e-print arXiv:1011.3771.

\bibitem{Ewings} R. A. Ewings \textit{et al.},
Phys. Rev. B \textbf{83}, 214519 (2011).

\bibitem{kaneshita} E. Kaneshita and T. Tohyama
Phys. Rev. B \textbf{82}, 094441 (2010).

\bibitem{Chuang} T.-M. Chuang \textit{et al.},
    Science \textbf{327} 181 (2010).

\bibitem{Dusza} A. Dusza \textit{et al.},
    EPL \textbf{93}, 37002 (2011).

\bibitem{Nakajima} M. Nakajima \textit{et al.},
J. Phys. Chem. Solids (to be published).


\bibitem{Wang}Q. Wang \textit{et al.},
    arXiv:1009.0271.

\bibitem{Kim}Y. K. Kim \textit{et al.},
    Phys. Rev. B \textbf{83}, 064509 (2011).

\bibitem{Tanatar} M.A. Tanatar \textit{et al.},
    Phys. Rev. B \textbf{81}, 184508 (2010).

\bibitem{Chu} J.-H. Chu \textit{et al.},
    Science \textbf{329}, 824 (2010).

\bibitem{Yi} M. Yi \textit{et al.},
 PNAS \textbf{108} 6878 (2011).

\bibitem{CCChen}C.-C. Chen, J. Maciejko, A. P. Sorini, B. Moritz, R. R. P. Singh, and T. P. Devereaux,
    Phys. Rev. B \textbf{82}, 100504(R) (2010).

\bibitem{Sugimoto} K. Sugimoto, E. Kaneshita, and T. Tohyama,
J. Phys. Soc. Jpn. \textbf{80}, 033706 (2011).

\bibitem{ZPYin}
Z. P. Yin, K. Haule and G. Kotliar, Nat. Phys. \textbf{7}, 294 (2011).

\bibitem{deGroot} F. M. F. de Groot, P. Kuiper, and G. A. Sawatzky,
Phys. Rev. B \textbf{57}, 14584 (1998).

\bibitem{veenendaal} M. van Veenendaal,
Phys. Rev. Lett. \textbf{96}, 117404 (2006).

\bibitem{ament} L. J. P. Ament, G. Ghiringhelli, M. M. Sala, L. Braicovich, and J. van den Brink,
Phys. Rev. Lett. \textbf{103}, 117003 (2009).

\bibitem{Ament11} L. J. P. Ament \textit{et al.},
Rev. Mod. Phys. \textbf{83}, 705 (2011).

\bibitem{Braicovich}L. Braicovich \textit{et al.},
     Phys. Rev. Lett. \textbf{104}, 077002 (2010).

\bibitem{chen} G. F. Chen, Z. Li, D. Wu, G. Li, W. Z. Hu, J. Dong, P. Zheng, J. L. Luo, and N. L. Wang,
    Phys. Rev. Lett. \textbf{100}, 247002 (2008)


\bibitem{j=1/2} For $j=1/2$, the spectral structure of $L_2$-edge RIXS is not so different; moreover, that of the spin excitation modes is the same in both cases.

\end{thebibliography}
\end{document}